# Machine Learning Forcefield for Silicate Glasses

Han Liu [a], Zipeng Fu [a,b], Yipeng Li [a], Nazreen Farina Ahmad Sabri [a], Mathieu Bauchy [a] *

[a] Physics of AmoRphous and Inorganic Solids Laboratory (PARISlab), Department of Civil and Environmental Engineering, University of California, Los Angeles, California, 90095, USA

[b] Department of Computer Science, University of California, Los Angeles, California, 90095, USA

## Abstract

Developing accurate, transferable, and computationally-efficient interatomic forcefields is key to facilitate the modeling of silicate glasses. However, the high number of forcefield parameters that need to be optimized render traditional parameterization methods poorly efficient or potentially subject to bias. Here, we present a new forcefield parameterization methodology based on ab initio molecular dynamics simulations, Gaussian process regression, and Bayesian optimization. By taking the example of glassy silica, we show that our methodology yields a new interatomic forcefield that offers an unprecedented description of the atomic structure of silica. This methodology offers a new route to efficiently parameterize new empirical interatomic forcefields for silicate glasses with very limited need for intuition.

## 1. Introduction

Classical molecular dynamics (MD) simulation is an effective tool to access the atomic structure of glass, which usually remains invisible from traditional experimental techniques [1–3]. In turn, better understanding the atomic structure of glasses is key to decipher their genome, that is, to understand how their composition and structure control their engineering properties [4]. However, the accuracy of glass modeling based on MD simulations largely depends on the reliability of the underlying empirical forcefield, i.e., the two-body (and sometimes three-body or more) interatomic potential [3]. Although *ab initio* molecular dynamics (AIMD) can, in theory, overcome these limitations, the high computational cost of this technique renders challenging glass simulations—which typically require large systems for statistical averaging and long timescales to slowly quench a melt down to the glassy state [3,5–7]. The development of new, improved empirical forcefields presently represents a bottleneck in glass modeling [8–10].

Empirical forcefields are typically based on functionals that depend on several parameters (e.g., partial atomic charges, etc.), which need to be properly optimized in order to minimize a given cost function [11–13]. One option is to define the cost function in terms of the difference between the structure or properties of the simulated system and available experimental data. However, such an optimization method may not yield a realistic forcefield in the case of glassy materials, since simulated and experimental glasses are prepared with significantly different cooling rate and, hence, their direct comparison may not be meaningful [5,14,15]. Although this problem can be partially overcome by conducting the optimization based on crystals rather than glasses, crystal-based potentials do not always properly describe the structure and properties of disordered, out-of-equilibrium glasses [13]. Alternatively, empirical forcefield can be parameterized based on AIMD simulations [9,16,17]. However, directly optimizing the forcefield in order to match with the interatomic forces or energy derived from AIMD sometimes results in unrealistic structures for the simulated glasses [9,12,18]. Recently, Kob, Huang et al. proposed a new optimization scheme, wherein the optimization cost function is defined based on the difference between the structure of a simulated liquid and that obtained by AIMD simulations in similar conditions [9,12,17]. However, such cost functions are very "rough," that is, they exhibit a large number of local minima (i.e., several sets of parameters yield similar, competitive results). This is a challenge as conventional gradient-based optimization methods (e.g., steepest descent or conjugate gradient) are highly inefficient to explore rough functions and are likely to yield a local minimum rather than the global one [19]. Due to this issue, conventional optimization methods are often biased, that is, their outcomes strongly depend on the starting point.

As an alternative route to conventional "intuition-based" forcefield parameterization, artificial intelligence and machine learning (ML) techniques have the potential to offer some efficient, non-biased optimization schemes [20,21]. To this end, several ML-based forcefields have been proposed [8,22,23]. However, although such forcefields can approach the accuracy of AIMD at a fraction of computing cost, their parametrization remains tedious and the complex form of the resulting forcefields render challenging their physical interpretation and their implementation [10,21,23–25]. For these reasons, ML-based forcefields have thus far mostly been limited to simple systems (e.g., comprising only one element at a time [10,26]), which does not yet offer a realistic path toward the simulation of complex multi-component glasses.

Here, we present a new less accurate, but more pragmatic approach to efficiently parametrize forcefields based on ML-based optimization. Our method is based on a predefined empirical potential form, wherein the parameters are optimized vs. AIMD simulations by Gaussian Process Regression and Bayesian optimization. We illustrate our new method by taking the example of glassy silica (g-$SiO_2$), an archetypal model for complex silicate glasses. Our method yields a new interatomic forcefield for g-$SiO_2$ that offers an unprecedented agreement with *ab initio* simulations. We demonstrate that, compared with traditional optimization methods, our ML-based optimization scheme is more efficient and non-biased. Overall, this work provides a realistic pathway toward the accurate, yet computationally efficient simulation of non-equilibrium disordered materials.

This paper is organized as follows. First, Sec. 2 describes the technical details of the simulations and parameterization strategy. The application of our method to glassy silica is then presented in Sec. 3. We then discuss the advantage of our approach over conventional optimization methods in Sec. 4. Finally, some conclusions are given in Sec. 5.

## 2. Methods
### 2.1 Reference ab initio simulations

A "reference" structure of a liquid silica system is first prepared by Car-Parrinello molecular dynamics (CPMD) [27]. The simulated system comprises 38 $SiO_2$ units (114 atoms) in a periodic cubic simulation box of length 11.982 Å—in accordance with the experimental density of 2.2 g/$cm^3$ [28]. The electronic structure is described with the framework of density functional theory and the choice of pseudopotentials for silicon and oxygen, exchange and correlation functions, and the plane-wave cutoff (70 Ry) are based on previous CPMD simulations of glassy silica [9,17]. A timestep of 0.0725 fs and a fictitious electronic mass of 600 atomic units are used. An initial liquid configuration is first prepared by conducting a classical MD run at 3600 K using the well-established van Beest–Kramer–van Santen (BKS) potential (see Sec. 2.2) [16]. The obtained configuration is then relaxed via CPMD at 3600 K for 3.5 ps at constant volume. Such duration is long enough considering the small relaxation time of the system at such elevated temperature. A subsequent dynamics of 16 ps is then used for statistical averaging and to compute the Si–Si, Si–O, and O–O partial pair distribution functions (PDFs) of the simulated liquid system. More details can be found in Ref. [9,17].

## 2.2 Classical molecular dynamics simulations

A new empirical forcefield for g-SiO$_2$ is then parameterized by conducting some classical MD simulations. The simulated system comprises 1000 SiO$_2$ units (3000 atoms) in a periodic cubic simulation box of length 35.661 Å, which corresponds to the experimental density of 2.2 g/cm$^3$ [28]. An initial configuration is first prepared by relaxing the system for 10 ps at 3600 K in the *NVT* ensemble. The partial PDFs of the simulated systems are then computed based on a subsequent *NVT* dynamics of 10 ps. A timestep of 1 fs is used for all simulations.

The interatomic potential energy between each pair of atom i, j is here described by adopting the Buckingham form [9,16]:

$$U_{ij} = \frac{q_i q_j}{4\pi\varepsilon_0 r_{ij}} + A_{ij}\exp\left(-\frac{r_{ij}}{\rho_{ij}}\right) - \frac{C_{ij}}{r_{ij}^6} + \frac{D_{ij}}{r_{ij}^{24}}$$  Eq. (1)

where $r_{ij}$ is the distance between each pair of atoms, $q_i$ is the partial charge of each atom ($q_O$ for oxygen, $q_{Si}$ for silicon, so that $q_O = -q_{Si}/2$), $\varepsilon_0$ is the dielectric constant, and the parameters $A_{ij}$, $\rho_{ij}$, $C_{ij}$, and $D_{ij}$ describe the short-range interactions. A cutoff of 8 Å is used for the short-range interactions. The long-range coulombic interactions are evaluated by damped shifted force (dsf) model [29] with a damping parameter of 0.25 and a cutoff of 8 Å. The last term serves as to add a strong repulsion at short distance to prevent the "Buckingham catastrophe" [9]. Since this term only aims to prevent any atomic overlap, the $D_{ij}$ parameters are not included in the present optimization and their value is fixed based on Ref. [9] (viz., $D_{ij}$ =113, 29, and 3423200 eV·Å$^{24}$ for O–O, Si–O, and Si–Si interactions, respectively). Note that this Buckingham formulation is chosen as it typically provides a good description of ionocovalent systems and has been shown to offer an improved description of g-SiO$_2$ as compared to alternative forms (e.g., Morse formulation) [12].

## 2.3 Optimization cost function

In total, the parametrization of this potential (Eq. 1) requires the optimization of 10 independent parameters, namely, the partial charge $q_{Si}$ and the short-range parameters {$A_{ij}$, $\rho_{ij}$, $C_{ij}$} for each of the three atomic pairs (Si–O, O–O, and Si–Si). This set of parameters is denoted Ξ thereafter. Following Kob and Huang *et al.*, we define the optimization cost function $R_\chi$ as follows [9,12,17]:

$$R_\chi = \sqrt{\frac{\chi_{SiO}^2 + \chi_{OO}^2 + \chi_{SiSi}^2}{3}}$$  Eq. (2)

where the $\chi_{\alpha\beta}^2$ terms capture the level of agreement between the partial PDFs obtained by classical MD and AIMD [30]:

$$\chi_{\alpha\beta}^2 = \frac{\sum_r [g_{\alpha\beta}^{AIMD}(r) - g_{\alpha\beta}^{MD}(r)]^2}{\sum_r [g_{\alpha\beta}^{AIMD}(r)]^2}$$  Eq. (3)

where $g_{\alpha\beta}^{\text{AIMD}}(r)$ and $g_{\alpha\beta}^{\text{MD}}(r)$ are the partial PDFs for each pair of atoms $\alpha$–$\beta$. Although other structural descriptors could be used to describe the structure of the simulated glasses, the PDF offers a convenient description of the short-range environment around each atom [9,17,31].

### 2.4 Forcefield optimization by machine learning

We now describe the ML-based optimization scheme used herein to parametrize the forcefield. An overview of the parametrization process is presented in Fig. 1. First, we create an initial dataset comprising some "known points," that is, the values of the cost function $R_\chi$ for select sets of parameters $\Xi$. Gaussian Process Regression (GPR) [32,33] is then used to interpolate the known points and assess the interpolation uncertainty over the entire parameter space. The Bayesian Optimization (BO) method [32] is then used to predict an optimal set of parameters $\Xi$ that offers the best "exploration vs. exploitation trade-off," that is, the best balance between (i) exploring the parameter space and reducing the model uncertainty and (ii) finding the global minimum of the cost function. The cost function $R_\chi\{\Xi\}$ associated with the set of parameters predicted by BO is subsequently calculated by conducting a classical MD simulation and comparing the structure of the simulated liquid with that of the reference AIMD configuration (see Sec. 2.3). This new datapoint $R_\chi\{\Xi\}$ is then added to the dataset. The new dataset is then used to refine the GPR-based interpolation and predict a new optimal set of parameters by BO. This cycle is iteratively repeated until a satisfactory minimum in the cost function is obtained. Finally, the global minimum predicted by BO is further refined by conducting a conjugate gradient (CG) optimization [19]. Each of these steps is further described in the following.

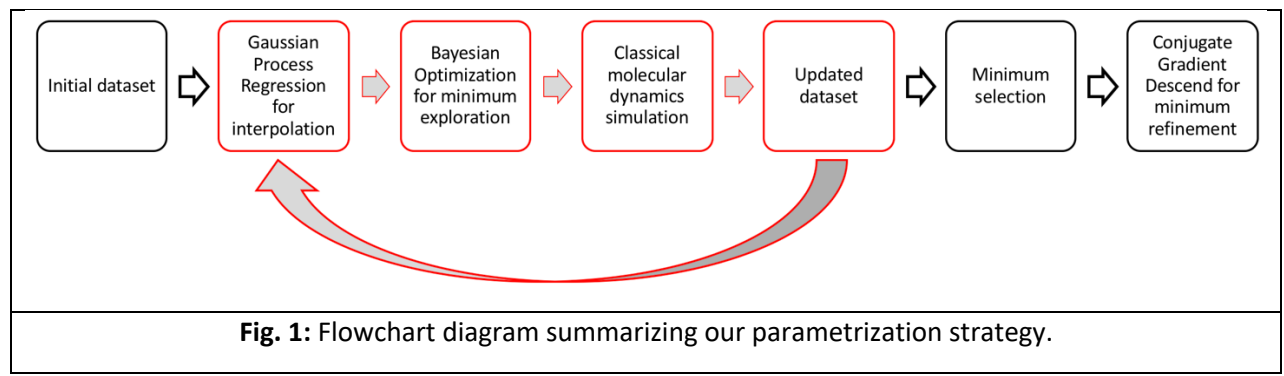

**Fig. 1:** Flowchart diagram summarizing our parametrization strategy.

*(i) Initial dataset*

As a starting point for our optimization method, we construct an initial dataset, which contains as inputs a selection of potential parameters $\Xi$ and as outputs the associated cost function $R_\chi$. Each of these datapoints is obtained by an independent MD simulation (see Secs. 2.2 and 2.3). This initial dataset offers an ensemble of known values for the cost function in 10-dimensional parameter space (i.e., for the 10 components in $\Xi$), which is used as a starting point for the iterative interpolation/exploration process described in the following. These initial values of $\Xi$ are chosen so as to uniformly span the targeted range of parameters. The initial dataset comprises about 1000 known points.

*(ii) Interpolation by Gaussian Process Regression*

The basic principle of GPR is to infer the (Gaussian-type) probability distribution of the values of the function that is interpolated based on a set of known points [32,33]. The interpolation process follows the following expression:

$$P\left(R_\chi(\Xi^*)\middle|\{R_\chi(\Xi_{\text{known}})\}\right) \Leftarrow$$

$$\begin{bmatrix} R_\chi(\Xi_1) \\ \vdots \\ R_\chi(\Xi_n) \\ R_\chi(\Xi^*) \end{bmatrix} \sim \text{Normal}\left(\begin{bmatrix} \mu_0(\Xi_1) \\ \vdots \\ \mu_0(\Xi_n) \\ \mu_0(\Xi^*) \end{bmatrix}, \begin{bmatrix} \Sigma_0(\Xi_1,\Xi_1) & \cdots & \Sigma_0(\Xi_1,\Xi_n) & \Sigma_0(\Xi_1,\Xi^*) \\ \vdots & \ddots & \vdots & \vdots \\ \Sigma_0(\Xi_n,\Xi_1) & \cdots & \Sigma_0(\Xi_n,\Xi_n) & \Sigma_0(\Xi_n,\Xi^*) \\ \Sigma_0(\Xi^*,\Xi_1) & \cdots & \Sigma_0(\Xi^*,\Xi_n) & \Sigma_0(\Xi^*,\Xi^*) \end{bmatrix}\right) \quad \text{Eq. (4)}$$

where $P\left(R_\chi(\Xi^*)\middle|\{R_\chi(\Xi_{\text{known}})\}\right)$ is the conditional probability of the value of the cost function $R_\chi$ for a given set of parameters $\Xi^*$ given the dataset of all the known points $\{R_\chi(\Xi_1), \; R_\chi(\Xi_2), \; \cdots \; R_\chi(\Xi_n)\}$, as denoted as $\{R_\chi(\Xi_{\text{known}})\}$. The conditional probability of $R_\chi(\Xi^*)$ is calculated using multivariate Gaussian distribution [34], where $\mu_0(\cdot)$ is the mean operation and $\Sigma_0(\cdot)$ is the covariance operation. There are many possible choices for the function-type of $\mu_0(\cdot)$ and $\Sigma_0(\cdot)$ and most can offer a reasonable extrapolation in the framework of multivariate Gaussian distribution [34]. Here, we adopt the Matern-type kernel for $\mu_0(\cdot)$ and $\Sigma_0(\cdot)$ [33,34]. In addition, to add some white-noise background during the interpolation [32], we also checked the intrinsic uncertainty of the cost function values yielded by the MD simulations by conducting a series of 10 independent MD simulations while keeping the same set of parameters $\Xi$ and calculating the standard deviation of the associated cost functions $R_\chi$. We find that the computed cost function values have a relative uncertainty of about 2% when $R_\chi < 100\%$ (i.e., for realistic forcefields) and can increase up to 10% for higher values of $R_\chi$ (i.e., for fairly unrealistic forcefields). This level of noise is not expected to significantly affect the shape of interpolation around the minimum positions of the cost function $R_\chi$.

Figure 2(a) shows an example of the outcome of a GPR-based interpolation. For illustration purposes, only the partial charge of the Si atoms $q_{\text{Si}}$ is here optimized, while the other 9 forcefield parameters are kept fixed and equal to those found in the original BKS potential [16]. A dataset comprising the values of the cost function $R_\chi$ for 5 values of $q_{\text{Si}}$ ranging from 1.6-to-3.2 is first constructed. The interpolated function and the uncertainty thereof (95% confidence interval) predicted by GPR is shown in Fig. 2(a). As expected, we observe that the interpolated function exhibits a minimum with respect to $q_{\text{Si}}$ (note that the $q_{\text{Si}}$ value used in the BKS potential is 2.4). Unsurprisingly, the uncertainty of the prediction is low at the vicinity of the known points and increases in between them.

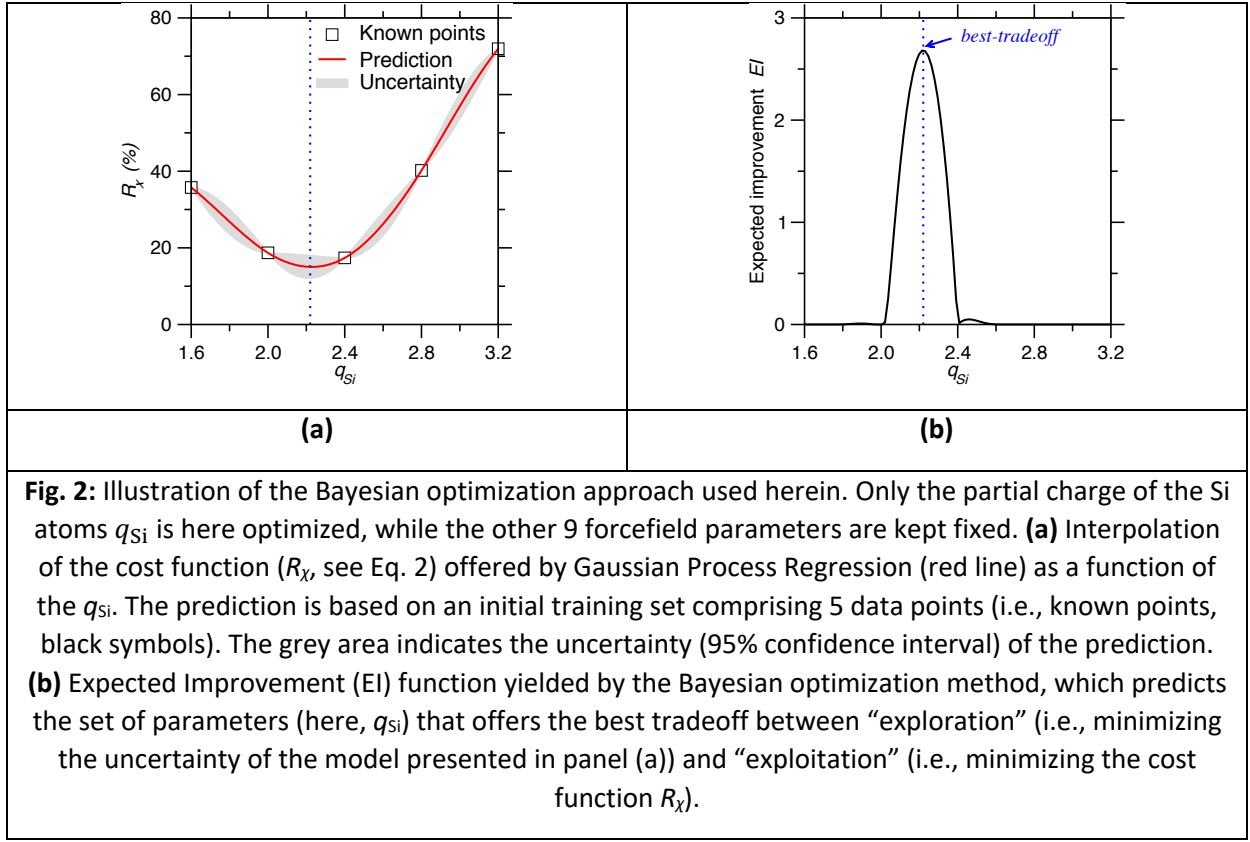

**Fig. 2:** Illustration of the Bayesian optimization approach used herein. Only the partial charge of the Si atoms $q_{Si}$ is here optimized, while the other 9 forcefield parameters are kept fixed. **(a)** Interpolation of the cost function ($R_\chi$, see Eq. 2) offered by Gaussian Process Regression (red line) as a function of the $q_{Si}$. The prediction is based on an initial training set comprising 5 data points (i.e., known points, black symbols). The grey area indicates the uncertainty (95% confidence interval) of the prediction. **(b)** Expected Improvement (EI) function yielded by the Bayesian optimization method, which predicts the set of parameters (here, $q_{Si}$) that offers the best tradeoff between "exploration" (i.e., minimizing the uncertainty of the model presented in panel (a)) and "exploitation" (i.e., minimizing the cost function $R_\chi$).

### (iii) Minimum exploration by Bayesian optimization

Based on the interpolated function $R_\chi(\Xi)$ and uncertainty $\sigma(\Xi)$ thereof predicted by GPR, the BO method is used to determine the next optimal set of parameters $\Xi$ to try based on an acquisition function that depends on $R_\chi(\Xi)$ and $\sigma(\Xi)$. Here, we adopt the expected improvement (EI) function, which is commonly used as acquisition function [32]:

$$EI(\Xi) = \begin{cases} [R_\chi(\hat{\Xi}) - R_\chi(\Xi)]\Phi(Z) + \sigma(\Xi)\phi(Z) & \text{if } \sigma(\Xi) > 0 \\ 0 & \text{if } \sigma(\Xi) = 0 \end{cases} \quad \text{Eq. (5)}$$

where $Z = [R_\chi(\hat{\Xi}) - R_\chi(\Xi)]/\sigma(\Xi)$, $R_\chi(\hat{\Xi})$ is the current minimum value of $R_\chi$ among all the known points (in other words, $\hat{\Xi}$ is the current optimal set of parameters), and $\Phi(Z)$ and $\phi(Z)$ are the cumulative distribution and probability density function of the standard normal distribution, respectively. By construction, the value of $EI(\Xi)$ is high (i) when the expected value of $R_\chi(\Xi)$ is smaller than the current best value $R_\chi(\hat{\Xi})$ or (ii) when the uncertainty $\sigma(\Xi)$ around the point $\hat{\Xi}$ is high. Therefore, the maximum position of $EI(\Xi)$ indicates either a point for which a better minimum position of $R_\chi$ than the current one is expected or a point belonging to a region of $R_\chi$ that has not been explored yet (i.e., $\sigma(\Xi)$ is high). Namely, the maximum position of $EI(\Xi)$ offers the best tradeoff between "exploration" (i.e., minimizing the uncertainty $\sigma(\Xi)$) and "exploitation" (i.e., minimizing the cost function $R_\chi(\Xi)$).

As an illustration of the BO approach, Fig. 2(b) shows the computed expected improvement function based on the interpolated function and uncertainty thereof shown in Fig. 2(a). As mentioned above, only the partial charge of the Si atoms $q_{Si}$ is here optimized, while the other 9 forcefield parameters are kept fixed and equal to those found in the original BKS potential [16]. As expected, we observe a noticeable maximum in the expected improvement function where the interpolated function $R_\chi$ is minimum (exploitation). Some secondary peaks are also observed in the high-uncertainty regions of the function in the vicinity of the minimum position.

### *(iv) Iterative refinement of the forcefield*

Finally, at each step of our iterative optimization scheme, the set of parameters $\Xi$ corresponding to the maximum of the expected improvement function is used to conduct an MD simulation and calculate the associated cost function value $R_\chi$. In turn, this new datapoint is added to the dataset. This enhances the accuracy of the GPR interpolation, which contributes to further refine the sampling of the cost function $R_\chi$ at the vicinity of its minimum positions. This iterative scheme is repeated until convergence is achieved, that is, until the cost function reaches a plateau and does not further decrease within 100 iterations.

This iterative refinement method is illustrated in Fig. 3. Here, for illustrative purposes, only two parameters ($q_{Si}$ and $A_{SiO}$) are optimized, while the other 8 forcefield parameters are kept fixed and equal to those found in the original BKS potential [16]. Figure 3(a) shows a contour plot of the cost function $R_\chi$ as a function of the two free parameters used in the optimization. We observe that, even in the case of only two free parameters, the cost function shows a rough dependence on the parameters and exhibits two distinct minima (i.e., the dark blue domains in Fig. 3(a)). Figure 3(a) also shows the pathway that is explored by the optimization algorithm in the ($q_{Si}$, $A_{SiO}$) space, that is, the set of parameters for which the expected improvement function is maximum after each step. We observe that the optimization quickly converges toward the global minimum of the cost function after only 5 iterations, after which the cost function $R_\chi$ shows a plateau around 10% (see Fig. 3(b)). This illustrates the efficiency of our optimization technique.

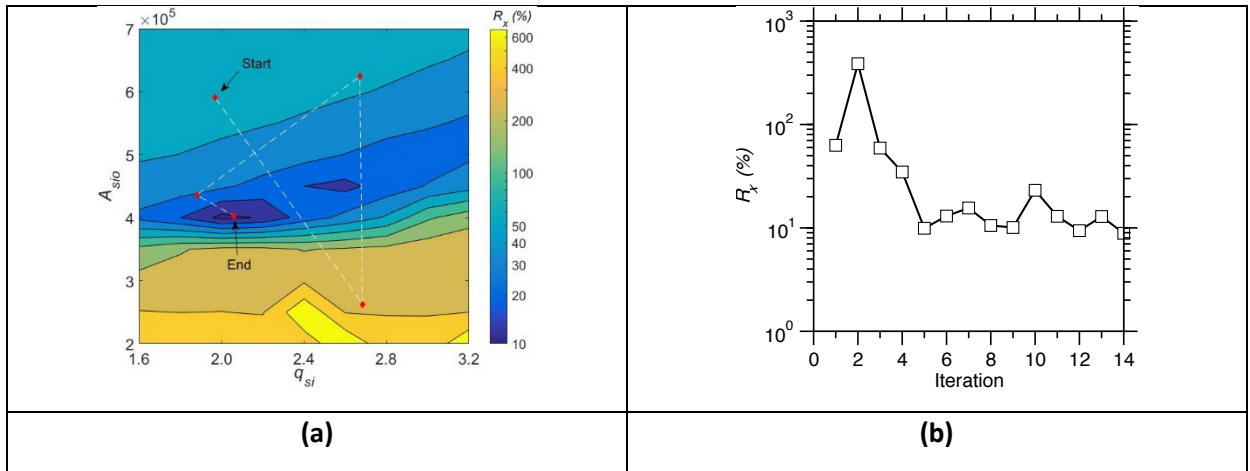

| (a) | (b) |

**Fig. 3:** Illustration of the iterative optimization approach used herein. Only the partial charge of the Si atoms $q_{Si}$ and the parameter $A_{SiO}$ are here optimized, while the other 8 forcefield parameters are kept fixed. **(a)** Contour plot showing the cost function $R_\chi$ as a function of $q_{Si}$ and $A_{SiO}$. The white dashed line indicates the path explored by the Bayesian optimization method until the global minimum in the cost function $R_\chi$ is identified. **(b)** Evolution of the cost function $R_\chi$ of the best-tradeoff position predicted by the Bayesian optimization during the optimization process.

## 2.5 Final refinement by conjugate gradient (CG)

Finally, the minimum identified by the iterative BO scheme is further refined by the CG method. Indeed, although the BO method can quickly identify the vicinity of the global minimum of a rough function, the CG method is more efficient to pinpoint the minimum position in a local basin of the cost function. Here, we adopt the nonlinear CG algorithm detailed in Ref [19]. In short, we first use the secant method to construct a quadratic interpolation of $R_\chi(\Xi)$ at the vicinity of the minimum identified by the iterative BO scheme and determine the new minimum predicted by the CG interpolation. We then repeat the quadratic construction (i.e., the linear search) around this new minimum position. This is used to approximate the minimum position of $R_\chi$ along the CG direction (i.e., the search direction). The maximum number of iterations of linear search in a search direction is set as 3. Then, starting from the identified new minimum position, we calculate the local gradient and find a new search direction based on Polak-Ribiere formula [19]. A new search direction is then determined from this starting point to identify a new minimum position. The iterative scheme is repeated until convergence, that is, when the new minimum position largely overlaps with the last minimum position, $R_\chi$ shows a plateau, and the squared sum of the local gradient converges toward zero and remains lower than the "zero" threshold (taken as 5 herein) within 10 iterations.

Figure 4 shows an illustration of the CG refinement step—starting from the minimum identified by the BO iterative scheme illustrated in Fig. 3. By exploring "downhill" the local minimum of the cost function (Fig. 4(a)), the CG allows us to further refine the position of the minimum—the cost function decreasing from 10% to about 9% (see Fig. 4(b)). As expected, the local gradient converges toward zero as the CG optimization proceeds (see Fig. 4(c)).

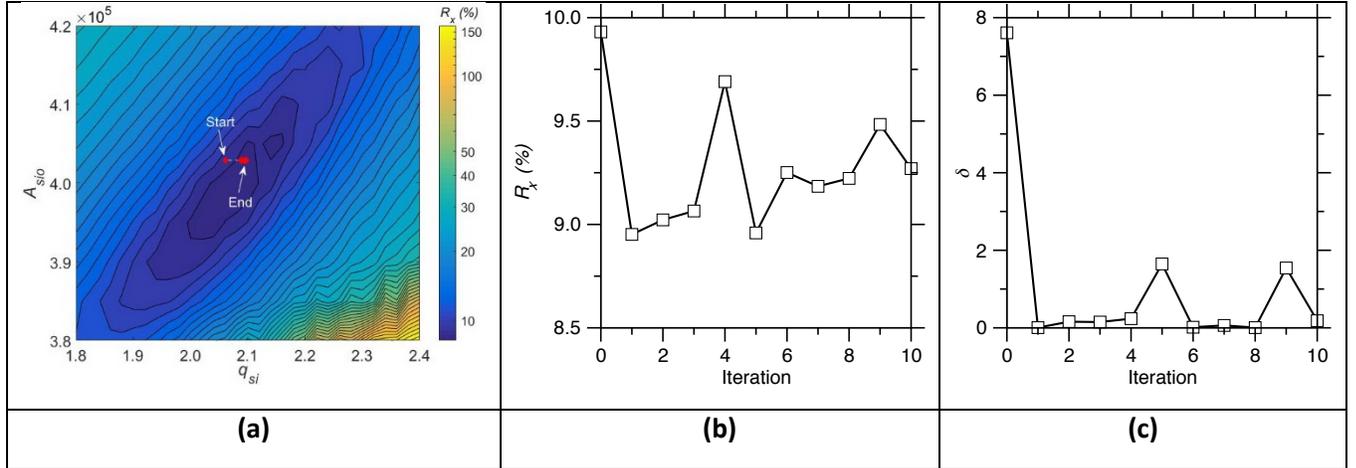

**Fig. 4:** Illustration of the final conjugate gradient optimization. Only the partial charge of the Si atoms $q_{Si}$ and the parameter $A_{SiO}$ are here optimized, while the other 8 forcefield parameters are kept fixed. **(a)** Contour plot showing the cost function $R_\chi$ as a function of $q_{Si}$ and $A_{SiO}$. The white dashed line indicates the path explored by the conjugate gradient optimization method until the minimum in the cost function $R_\chi$ is identified. **(b)** Evolution of the cost function $R_\chi$ during the conjugate gradient optimization process. **(c)** Evolution of the squared-sum of the local gradient $\delta$ during the conjugate gradient optimization process.

# 3. Results

## 3.1 New interatomic forcefield for glassy silica

We first conduct an optimization of the force-field while keeping the Si–Si interaction energy term as being zero (that is, $A_{\text{SiSi}} = C_{\text{SiSi}} = 0$ and $\rho_{\text{SiSi}} = 1$ Å). This choice is motivated by the fact that the original BKS potential does not comprise any Si–Si interaction energy terms, which suggests that the optimization of these terms may not be necessary. Decreasing the number of variable parameters allows us to increase the efficiency of the optimization. In addition, decreasing the complexity of the forcefield decreases the risk of overfitting, which, in turn, is likely to increase the transferability of the new forcefield to new systems that are not considered during its training.

The forcefield parameters obtained after the BO and CG optimization steps are listed in Tab. 1. The performance of our forcefield (as quantified in terms of the final cost function $R_\chi$) is compared with that of select alternative potentials in Tab. 2. We find that our new ML forcefield yields a $R_\chi$ of 8.77%. This constitutes a significant improvement with respect to the well-established BKS potential (for which $R_\chi$ is about 17%) [16,35]. Our new potential is also found to be slightly better than the CHIK potential parameterized by Kob *et al.* [17]. This is not surprising as the CHIK potential was obtained based on the optimization of a slightly different cost function [17]. However, it is worth noting that our new potential exhibits a lower complexity than the CHIK parametrization (which comprises 3 extra parameters for the Si–Si interactions). The role played by these extra parameters is discussed in Sec. 4.

In details, we find that the parameters of our ML forcefield are significantly different from those of the original BKS potential—which illustrates the roughness of the cost function. Interestingly, we find that our ML potential relies on a partial charge for Si atoms that is significantly smaller than that of the BKS potential (+1.955 vs. +2.4 for BKS). In turn, this value is close to that of the CHIK (+1.91 [17]) and Wang–Bauchy potential (+1.89 [13]). This suggests that "soft potentials" (i.e., which relies on lower partial charges) appear to consistently perform better than the stiffer ones, e.g., BKS. The cost function $R_\chi$ associated with each interatomic pair (see Eq. 3) is also provided in Tab. 2. Overall, we note that our ML potential consistently offers an improved description of the interatomic structural order for each pair of atoms.

**Tab. 1:** Parameters of our new interatomic potential "ML" (see Eq. 1). The partial charges are indicated as subscripts for each pair of atoms.

| Atomic pairs | $A$ (eV) | $\rho$ (Å) | $C$ (eV·Å$^6$) |
|---|---|---|---|
| Si$^{+1.955}$ – O$^{-0.9775}$ | 20453.601 | 0.191735 | 93.496 |
| O$^{-0.9775}$ – O$^{-0.9775}$ | 1003.387 | 0.356855 | 81.491 |
| Si$^{+1.955}$ – Si$^{+1.955}$ | 0 | 1 | 0 |

**Tab. 2:** Comparison of our new "ML" forcefield with select alternative classical potentials, namely, "BKS" [16] and "CHIK" [17].

| Forcefield | $R_\chi^{SiO}$ (%) | $R_\chi^{OO}$ (%) | $R_\chi^{SiSi}$ (%) | Global $R_\chi$ (%) |
|---|---|---|---|---|
| ML | 7.35 | 3.58 | 12.80 | 8.77 $\pm$ 0.25 |
| BKS | 21.45 | 12.90 | 15.54 | 17.01 $\pm$ 0.25 |
| CHIK | 12.29 | 6.09 | 11.76 | 10.43 $\pm$ 0.25 |

### 3.2 Partial pair distribution functions

We now further analyze the structure of the simulated $SiO_2$ liquid (i.e., at 3600 K). Figure 5 shows the partial PDFs predicted by our new ML forcefield. The data are compared with the reference *ab initio* partial PDFs used for the training of the potential [17] as well as those predicted by the BKS potential [16]. Overall, we find that our ML forcefield offers an excellent agreement with AIMD simulations—although this is not surprising as our forcefield is specifically trained to match these data. Nevertheless, these results show that the Buckingham formulation adopted herein is appropriate for the $SiO_2$ system and further supports the ability of our optimization method to offer a robust parametrization. We note that the average Si–Si distance predicted by our potential is slightly shifted with respect to that obtained in AIMD simulations (see Fig. 5(c)). As discussed in Sec. 4.3, this may arise from a general limitation of the Buckingham formulation. Nevertheless, our ML forcefield offers a significant improvement with respect to the BKS potential, especially in the case of the Si–O and O–O partial PDFs (see also Tab 2). We note that our ML forcefield systematically predicts some PDF peaks that are broader than those predicted by BKS, which suggests that our forcefield yields a slightly more disordered structure. This may be linked with the fact that our potential relies on lower partial charge values (i.e., softer Coulombic interactions).

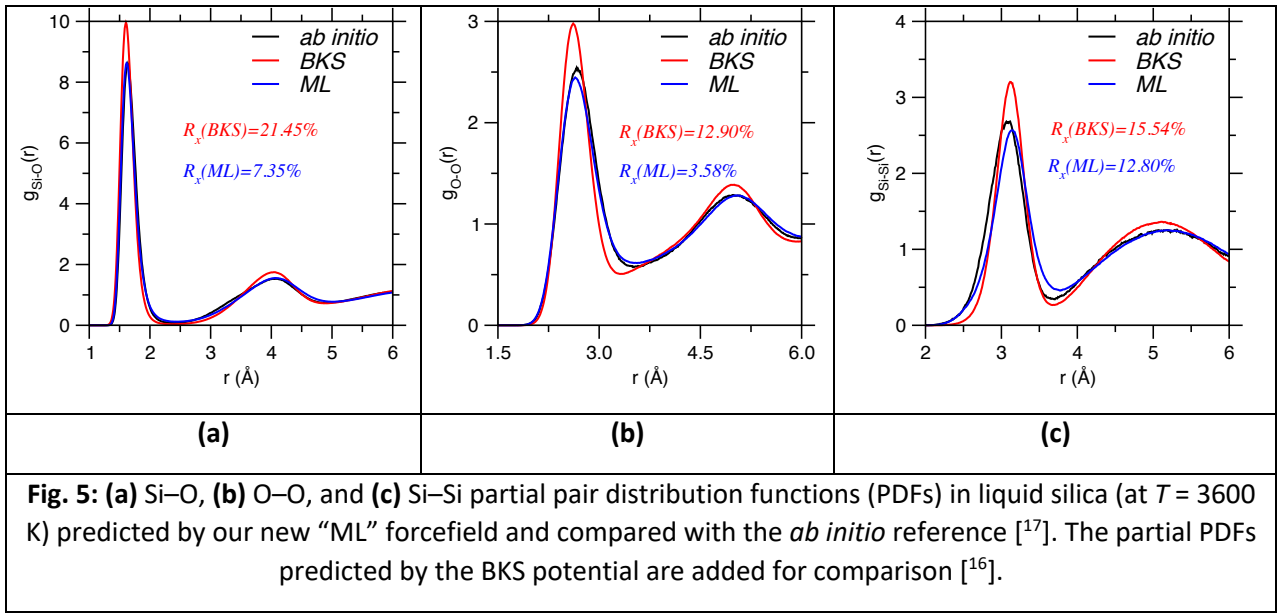

**Fig. 5: (a)** Si–O, **(b)** O–O, and **(c)** Si–Si partial pair distribution functions (PDFs) in liquid silica (at *T* = 3600 K) predicted by our new "ML" forcefield and compared with the *ab initio* reference [17]. The partial PDFs predicted by the BKS potential are added for comparison [16].

### 3.3 Partial bond angle distributions

We now focus on the angular environment around each atom. To this end, Fig. 6 shows the O–Si–O and Si–O–Si partial bond angle distributions (PBADs) predicted by our ML forcefield for the liquid silica system (at $T$ = 3600 K). The data are compared with those obtained by *ab initio* simulations [17] and predicted by the BKS potential [16]. Overall, we observe that the PBADs predicted by our ML forcefield are in very good agreement with those obtained by *ab initio* simulations—with a significant improvement with respect to the BKS potential. This is significant as the PBADs are not explicitly included in the cost function used herein and such 3-body correlations are not fully encoded in 2-body correlations (i.e., as captured by the partial PDFs). As such, these results offer a strong a posteriori validation of the performance of our new ML forcefield.

As expected, our forcefield yields a tetrahedral environment for Si atoms (with an average O–Si–O angle of about 109°). However, we note that the O–Si–O PBAD predicted by our ML forcefield is broader than that obtained with BKS, which suggests that our potential yields a slightly more disordered angular environment around Si atoms. Again, this may be linked with the fact that our potential relies on lower fictive charges than BKS (see Sec. 3.2). In contrast, we observe that our forcefield slightly overestimates the value of Si–O–Si angle with respect to AIMD simulations. This is likely linked with the fact that our potential overestimates the Si–Si average distance (see Sec. 3.2), which appears to be a general limitation of the two-body Buckingham formulation. Nevertheless, the Si–O–Si PBAD yielded by our forcefield is significantly improved with that obtained by BKS (which tends to largely overestimate the average Si–O–Si angle).

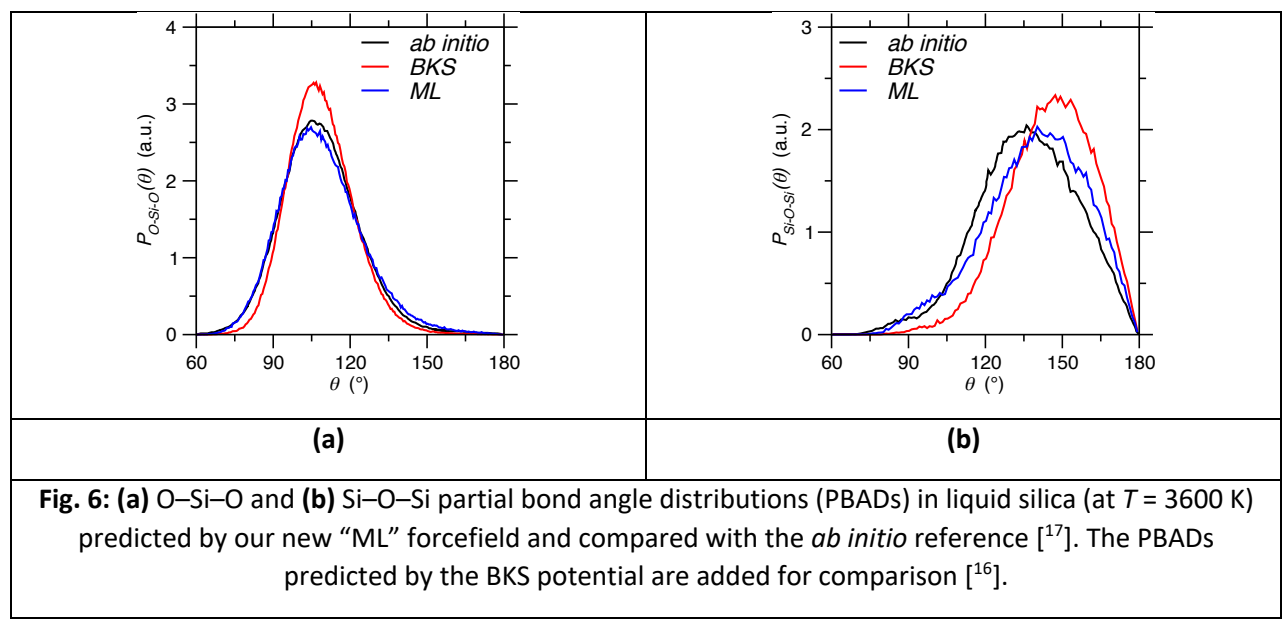

(a)      (b)

**Fig. 6: (a)** O–Si–O and **(b)** Si–O–Si partial bond angle distributions (PBADs) in liquid silica (at $T$ = 3600 K) predicted by our new "ML" forcefield and compared with the *ab initio* reference [17]. The PBADs predicted by the BKS potential are added for comparison [16].

## 4. Discussion
### 4.1 Comparison between gradient-based and machine-learning-based optimization

We now discuss the performance of our ML-based optimization method by comparing its ability to identify the global minimum of the cost function with that of the conjugate gradient method. Here, for illustrative purposes, only two parameters ($q_{Si}$ and $A_{SiO}$) are optimized in both cases, while the other 8 forcefield parameters are kept fixed and equal to those found in the original BKS potential [16]. As shown in Fig. 7(a), the cost function $R_\chi$ shows a very rough dependence on the forcefield parameters—wherein the level of roughness appears to increase when upon zooming on the fine details of the landscape (see Fig. 7(b)). The pathways explored (starting from the same initial point) upon the ML-based and CG-based optimizations in the ($q_{Si}$, $A_{SiO}$) space is shown in Fig. 7(a). We observe that, in contrast with our ML optimization method, the CG optimization quickly get "stuck" in a local minimum of the cost function (see Fig. 7(c)) and does not succeed at identifying the global minimum. This highlights the fact that traditional gradient-based optimization methods are not appropriate in the case of such high-roughness function and, hence, are highly biased based on the chosen starting point. Although the efficiency of the CG method could certainly be improved by adjusting some parameters (e.g., the learning rate and step length [19]), such fine-tuning necessarily requires some level of intuition or trial-and-error optimization, which is a clear advantage of the present ML approach.

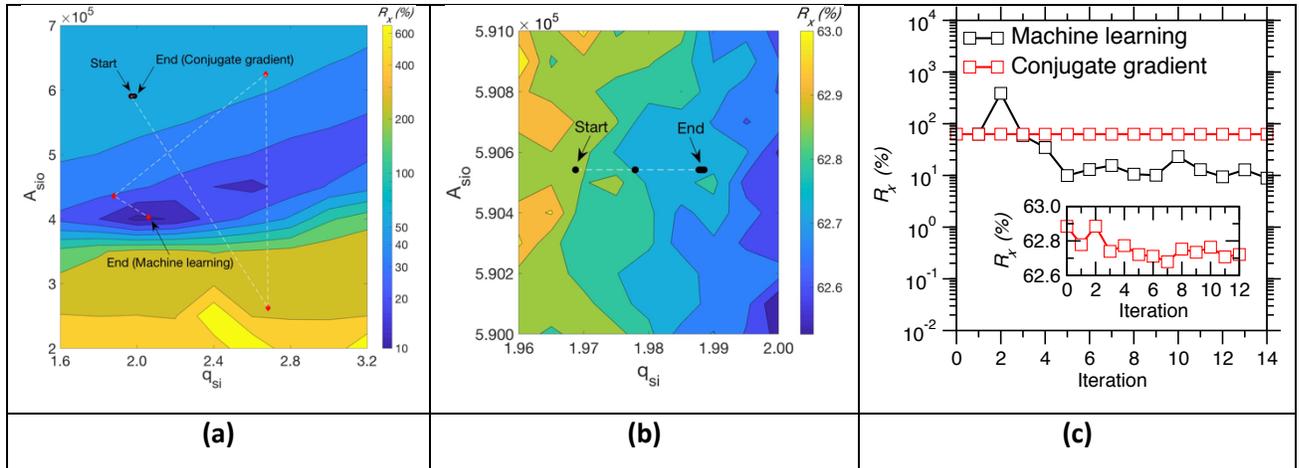

**Fig. 7:** Comparison between machine learning and conjugate gradient optimization. Only the partial charge of the Si atoms $q_{Si}$ and the parameter $A_{SiO}$ are here optimized, while the other 8 forcefield parameters are kept fixed. **(a)** Contour plot showing the cost function $R_\chi$ as a function of $q_{Si}$ and $A_{SiO}$. The red and black circles indicate the path explored upon machine learning and conjugate gradient optimization, respectively. Panel **(b)** is a zoom of the data presented in panel (a) to better observe the path explored upon conjugate gradient optimization. **(c)** Evolution of the cost function $R_\chi$ during the machine learning and conjugate gradient optimizations. The inset is a zoom on data obtained in the case of conjugate gradient optimization.

### 4.2 Lessons from the BKS potential

It worth further focusing on the BKS potential [16] to establish some general conclusions regarding the development of interatomic forcefields for glassy materials. The well-established BKS potential was parameterized by sequentially optimizing the O–O and Si–O energy terms so as to an isolated SiO$_4$ cluster (saturated by 4 H atoms) matches with *ab initio* simulations. Si–Si energy terms were forced to be zero. In addition, the experimental elastic constants of silica were used to discriminate several competing sets of parameters. This suggests that the BKS potential is specifically trained to offer an excellent description of the interatomic potential in the vicinity of its equilibrium position. In details, the position of the minima of each energy terms is encoded in the geometry of the isolated $SiO_4$ cluster (i.e., the average interatomic distances), while the curvature of the potential energy at the vicinity of the equilibrium position is encoded in the elastic constant. Nevertheless, as detailed in Sec. 3, this optimization scheme tends to overestimate the radial and angular order around Si atoms (see Figs. 5 and 6). This suggests that optimization schemes placing a strong emphasis on describing the shape of the forcefield in the very close vicinity of the equilibrium position may not be appropriate to describe the disordered structure of glasses, which are intrinsically out-of-equilibrium and wherein the atoms are not exactly located at their minimum-energy positions. For instance, the degree of asymmetry of the forcefield is likely to play a key role in governing the structure of disordered materials and may not be efficiently trained by considering only equilibrium structures (e.g., crystals or isolated clusters). This suggests that parametrization methods based on liquid structures (as the present one) may be more appropriate to develop new improved forcefields for complex glasses.

### 4.3 Effect of the forcefield complexity and risk of overfitting

Finally, we discuss the effect of the complexity (i.e., number of parameters) of the forcefield. To this end, we repeat the optimization process previously described while (i) excluding O–O interactions, i.e., including only Si–O energy terms (referred to as ML–SiO potential thereafter) and (ii) adding Si–Si interactions, i.e., in addition to the Si–O and O–O terms (referred to as ML–ALL potential thereafter). In order of increasing complexity, the three potentials comprise 4, 7, and 10 variable parameters, respectively (i.e., 3 parameters per interatomic pair and the Si partial charge). These extra parametrizations aim to investigate (i) whether O–O interaction terms are truly necessary to predict a realistic structure for glassy silica and (ii) the extent to which incorporating Si–Si energy terms could improve the performance of our ML forcefield. More generally, this analysis is conducted to ensure that our forcefield relies on the right level of complexity, that is, to confirm that our model is not under- or over-fitted.

**Tab. 3:** Parameters of the interatomic potential "ML-SiO" (which only considers Si–O interactions). The partial charges are indicated as subscripts for each pair of atoms.

| Atomic pairs | $A$ (eV) | $\rho$ (Å) | $C$ (eV·Å$^6$) |
| --- | --- | --- | --- |
| $Si^{+1.484} - O^{-0.742}$ | 3968.508 | 0.187600 | 0.677 |
| $O^{-0.742} - O^{-0.742}$ | 0 | 1 | 0 |
| $Si^{+1.484} - Si^{+1.484}$ | 0 | 1 | 0 |

**Tab. 4:** Parameters of the interatomic potential "ML-ALL" (which includes Si–Si interactions). The partial charges are indicated as subscripts for each pair of atoms.

| Atomic pairs | $A$ (eV) | $\rho$ (Å) | $C$ (eV·Å$^6$) |
|---|---|---|---|
| Si$^{+1.955}$ – O$^{-0.9775}$ | 20453.601 | 0.191735 | 93.496 |
| O$^{-0.9775}$ – O$^{-0.9775}$ | 1003.387 | 0.356855 | 81.491 |
| Si$^{+1.955}$ – Si$^{+1.955}$ | 2643.111 | 0.303616 | 232.009 |

The parameters obtained for the ML-SiO and ML-ALL potentials are listed in Tab. 3 and Tab. 4, respectively, while Fig. 8 presents the final cost function $R_\chi$ for each potential. We find that the parameters of the ML-SiO potential significantly differ from those of the ML potential described in Sec. 3. In particular, we obtain a very small Si partial charge of +1.484. As shown in Fig. 8, the ML-SiO potential offers a very description of the structure of silica (i.e., high final $R_\chi$ value). This confirms that, as expected, O–O interactions play a key role to predict a realistic SiO$_2$ structure and that the ML-SiO model is clearly under-fitted. In contrast, we note that the parameters of the ML-ALL forcefield are largely similar to those of the ML potential described in Sec. 3. As shown in Fig. 8, the ML-ALL potential offers a slight improvement in the description of the structure of silica, which manifests itself by a slight decrease in $R_\chi$ from 8.77% to 7.20%. Although this improvement is higher than the level of uncertainty in the $R_\chi$ values, it remains small as compared to the difference between the ML and ML-SiO forcefields. This suggests that Si–Si interactions only play a minor role in controlling the structure of silica. In turn, this small improvement comes with a significantly higher degree of complexity (i.e., 3 extra parameters), which suggests that the ML-ALL potential may be overfitted.

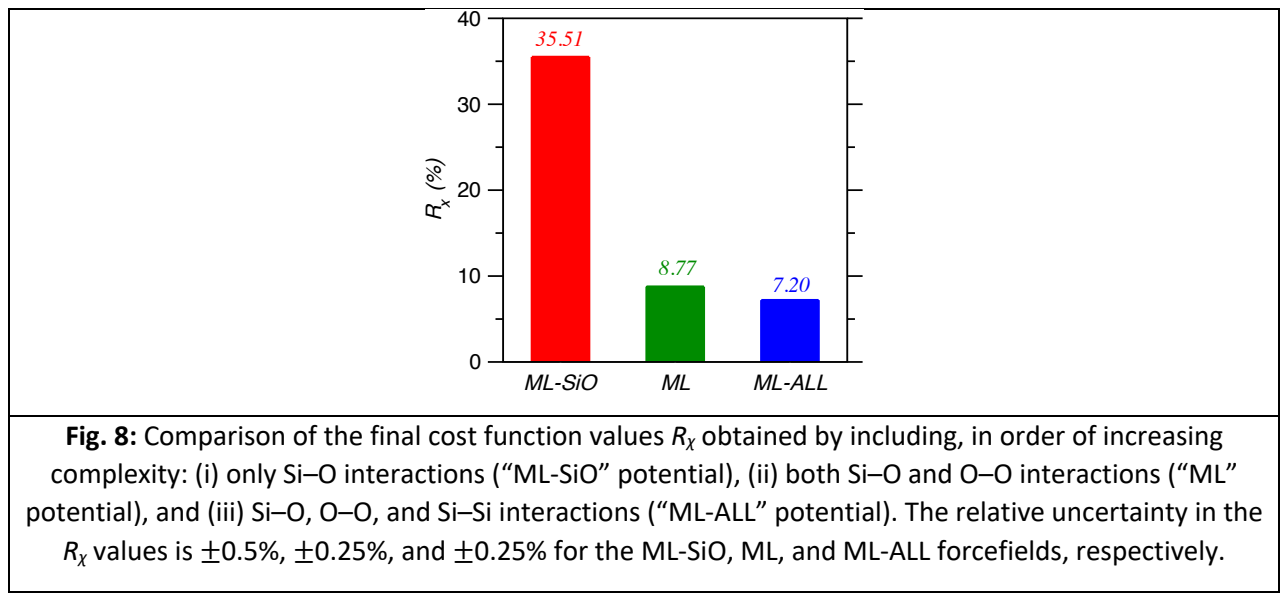

**Fig. 8:** Comparison of the final cost function values $R_\chi$ obtained by including, in order of increasing complexity: (i) only Si–O interactions ("ML-SiO" potential), (ii) both Si–O and O–O interactions ("ML" potential), and (iii) Si–O, O–O, and Si–Si interactions ("ML-ALL" potential). The relative uncertainty in the $R_\chi$ values is ±0.5%, ±0.25%, and ±0.25% for the ML-SiO, ML, and ML-ALL forcefields, respectively.

Finally, we further investigate the effect of the complexity of the forcefield on the structure of the simulated liquid silica system. To this end, Figure 9 shows a comparison of the partial PDFs obtained with each potential. Overall, we note that, in the absence of any O–O interactions (i.e., with the ML-SiO potential), the simulated system exhibits a very unrealistic structure. Although this forcefield succeeds at predicting a reasonable average Si–O average interatomic distance (see Fig. 9(a)), it completely fails to properly model O–O correlations (see Fig. 9(b)). This confirms that including the O–O interactions is necessary to properly describe the tetrahedral structure of Si atoms. In contrast, we note that the structure predicted by the ML-ALL forcefield is largely similar to that offered by the ML potential, which confirms that Si–Si interactions play a fairly trivial role. Although we observe that taking into account Si–Si interactions offers a slight improvement in the Si–Si partial PDF, the average Si–Si distance remains overestimated with respect to that predicted by AIMD. This further suggests that this discrepancy is an intrinsic limitation of the Buckingham formulation used herein. Although the inclusion of 3-body energy terms could overcome this limitation, this would come with a significant increase in computing cost and model complexity. Overall, these results confirm that the ML parametrization presented in Tab. 1 yields an excellent description of the structure of silica and offers the best balance between accuracy and complexity.

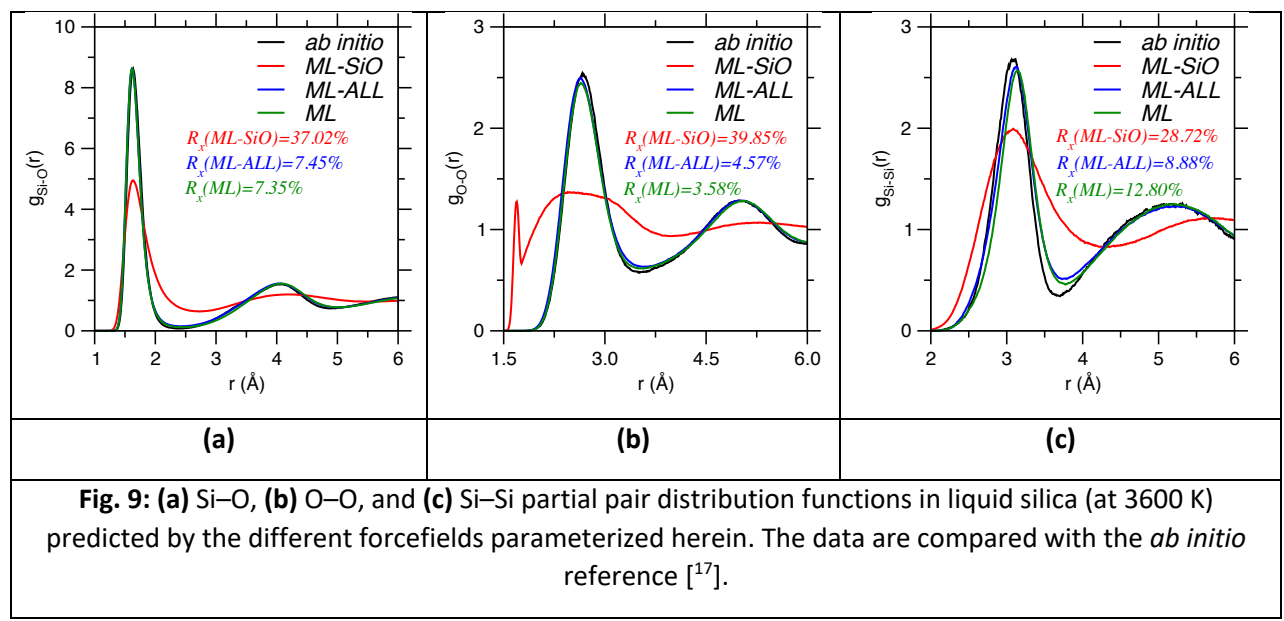

**Fig. 9: (a)** Si–O, **(b)** O–O, and **(c)** Si–Si partial pair distribution functions in liquid silica (at 3600 K) predicted by the different forcefields parameterized herein. The data are compared with the *ab initio* reference [17].

## 5. Conclusions

Overall, this study establishes a general and versatile framework to accelerate the parametrization of new, improved empirical forcefields for disordered materials. As shown herein with the example of silica, our method makes it possible to quickly reoptimize previous well-established potentials (e.g., the BKS forcefield). By using as a reference some liquid structures prepared by *ab initio* molecular dynamics simulations, our parametrization scheme is better suited for glass modeling than alternative methods based on equilibrium crystal or isolated atomic clusters. Importantly, the use of machine learning rather than alternative traditional optimization methods (e.g., conjugate gradient) (i) drastically improves the efficiency of the parametrization procedure, (ii) suppresses the risk of bias resulting from arbitrary choices regarding the starting point of the optimization, and (iii) significantly reduces the role played by "personal intuition" during the parametrization. As a key advantage over alternative conventional method, the present ML-based parametrization method is highly scalable and, hence, can be used to parametrize multi-component systems (i.e., many forcefield parameters can be optimized simultaneously). Overall, this work establishes an efficient, pragmatic method to develop new improved forcefields for the simulation of complex "real-world" materials—which addresses an immediate concern since more accurate ML-based forcefields that do rely on a predefined functional are unlikely to be available for complex multi-component systems in the near future.

## Acknowledgments

This work was supported by the National Science Foundation under Grants No. 1562066, 1762292, and 1826420.